\documentclass[aip,amsmath,amssymb,reprint,floatfix]{revtex4-1}

\usepackage{graphicx} 
\usepackage{dcolumn}
\usepackage{bm}
\usepackage[utf8]{inputenc}
\usepackage[T1]{fontenc}
\usepackage{mathptmx}
\usepackage{etoolbox}

\makeatletter
\def\@email#1#2{%
 \endgroup
 \patchcmd{\titleblock@produce}
  {\frontmatter@RRAPformat}
  {\frontmatter@RRAPformat{\produce@RRAP{*#1\href{mailto:#2}{#2}}}\frontmatter@RRAPformat}
  {}{}
}%
\makeatother

\usepackage{xspace}              
\usepackage{tikz}                
\usetikzlibrary{bayesnet}        
\usepackage{siunitx}             
\usepackage[version=3]{mhchem} 
\ifdefined\unit\else             
  \ifdefined\NewCommandCopy
    \NewCommandCopy\unit\si
  \else
    \NewDocumentCommand\unit{O{}m}{{\,}\si[#1]{#2}}
  \fi
\fi

\DeclareSIUnit\angstrom{\text{\AA}}
\DeclareSIUnit\atm{\text{atm}}
\DeclareSIUnit\atmosphere{\text{atm}}

\begin{document}
\preprint{AIP/123-QED}

\newcommand{\mOH}{\ensuremath{\cdot\mathrm{OH}}\xspace}
\newcommand{\radH}{\ensuremath{\cdot\mathrm{H}}\xspace}
\newcommand{\mHTO}{\ensuremath{\mathrm{H}_2\mathrm{O}}\xspace}
\newcommand{\posHTO}{\ensuremath{\mathrm{H}_2\mathrm{O}^+}\xspace}
\newcommand{\mHP}{\ensuremath{\mathrm{H}^{+}}\xspace}
\newcommand{\mH}{\ensuremath{\mathrm{H}^{+}}\xspace}
\newcommand{\DiOH}{\ensuremath{D_{\cdot\mathrm{OH}}}\xspace}
\newcommand{\DiHTO}{\ensuremath{D_{\mathrm{H}_2\mathrm{O}}}\xspace}
\newcommand{\DiH}{\ensuremath{D_{\mH}}\xspace}
\newcommand{\DiP}{\ensuremath{D_{\mathrm{P}}}\xspace}
\newcommand{\DiHP}{\ensuremath{D_{\mHP}}\xspace}
\newcommand{\EO}{\ensuremath{\epsilon_{\mathrm{O}}}\xspace}
\newcommand{\EH}{\ensuremath{\epsilon_{\mathrm{H}}}\xspace}
\newcommand{\RO}{\ensuremath{r_{\min\mathrm{O}}{/ \scriptstyle 2}}\xspace}
\newcommand{\RH}{\ensuremath{r_{\min\mathrm{H}}{/ \scriptstyle 2}}\xspace}
\newcommand{\RA}{\ensuremath{r_{\min A}{/ \scriptstyle 2}}\xspace}
\newcommand{\RB}{\ensuremath{r_{\min B}{/ \scriptstyle 2}}\xspace}
\newcommand{\muOH}{\ensuremath{\mu_{\cdot\mathrm{OH}}}\xspace}
\newcommand{\muHTO}{\ensuremath{\mu_{\mathrm{H}_2\mathrm{O}}}\xspace}
\newcommand{\gammaOH}{\ensuremath{\gamma_{\cdot\mathrm{OH}}}\xspace}
\newcommand{\gammaHTO}{\ensuremath{\gamma_{\mathrm{H}_2\mathrm{O}}}\xspace}
\newcommand{\DMDOH}{\ensuremath{D_{\mathrm{MD}}^{\cdot\mathrm{OH}}}\xspace}
\newcommand{\DMDHTO}{\ensuremath{D_{\mathrm{MD}}^{\mathrm{H}_2\mathrm{O}}}\xspace}
\newcommand{\SMDOH}{\ensuremath{S_{\mathrm{MD}}^{\cdot\mathrm{OH}}}\xspace}
\newcommand{\SMDHTO}{\ensuremath{S_{\mathrm{MD}}^{\mathrm{H}_2\mathrm{O}}}\xspace}
\newcommand{\tauOH}{\ensuremath{\tau_{\cdot\mathrm{OH}}}\xspace}
\newcommand{\tauHTO}{\ensuremath{\tau_{\mathrm{H}_2\mathrm{O}}}\xspace}
\newcommand{\RcH}{\ensuremath{R_{\mH}}\xspace}
\newcommand{\RcP}{\ensuremath{R_{\mathrm{P}}}\xspace}
\newcommand{\rOH}{\ensuremath{r_{\mOH}}\xspace}
\newcommand{\rH}{\ensuremath{r_{\mH}}\xspace}
\newcommand{\rP}{\ensuremath{r_{\mathrm{P}}}\xspace}
\newcommand{\e}{\ensuremath{e^-_{\mathrm{aq}}}\xspace}

\title[OH Radical Diffusion Coefficient at Low Temperature and High Pressure]{Bayesian Estimation of the Hydroxyl Radical Diffusion Coefficient at Low Temperature and High Pressure from Atomistic Molecular Dynamics
}

\author{Carter T. Butts}
\email{buttsc@uci.edu, rwmartin@uci.edu}
\affiliation{Departments of Sociology, Statistics, Computer Science, and EECS, University of California, Irvine}
\author{Rachel W. Martin}
\affiliation{Departments of Chemistry and Molecular Biology and Biochemistry; University of California, Irvine}
\date{7/28/21}

\begin{abstract}
The hydroxyl radical is the primary reactive oxygen species produced by the radiolysis of water, and is a significant source of radiation damage to living organisms.  Mobility of the hydroxyl radical at low temperatures and/or high pressures is hence a potentially important factor in determining the challenges facing psychrophilic and/or barophilic organisms in high-radiation environments (e.g., ice-interface or undersea environments in which radiative heating is a potential heat and energy source).  Here, we estimate the diffusion coefficient for the hydroxyl radical in aqueous solution, using a hierarchical Bayesian model based on atomistic molecular dynamics trajectories in TIP4P/2005 water over a range of temperatures and pressures.
\end{abstract}
\maketitle


Ionizing radiation is a feature of both terrestrial and extraterrestrial environments, presenting challenges as well as opportunities for living organisms.  On one hand, radiation can damage biological molecules either directly or via the production of reactive chemical species that modify or degrade them \citep{Blanco:2018tp}.  On the other, ionizing radiation can act as a non-photosynthetic energy source for microbial communities, in some cases producing chemical species suitable for chemotrophy \citep{Onstott:2003wz} or in other cases maintaining a habitable environment by temperature elevation and/or maintenance of liquid water pockets via ice melting \citep{Tarnas:2018vo, Ojha:2021uu, Tarnas:2021wn}.  While some terrestrial organisms are known to tolerate high levels of radiation either transiently or on an ambient basis \citep{White:1999ux, Cavicchioli:2002wg, Munteanu:2015ws}, the role of ionizing radiation in determining habitability in a broader biological context remains largely open.

For conventional microbial organisms in aqueous environments, the predominant threat posed by ionizing radiation is the formation of reactive oxygen species due to the radiolysis of water \citep{LaVerne:2000wz}.  Many species are produced, including hydrogen peroxide,  the superoxide radical, and the hydroperoxyl radical; \citep{Matheson:1964wm} however, the hydroxyl radical (\mOH) is the dominant source of prompt radiation damage for solvated biomolecules due to its high production rate, reactivity, and unsuitability for enzymatic processing \citep{Ghosal:2005we}.  In terrestrial mesophilic organisms under typical cellular conditions, \mOH survives on sufficiently long time scales to diffuse to and damage biological macromolecules \citep{roots.okada:rr:1975}, prominently including proteins and DNA.  Although the problems associated with DNA damage are well-appreciated, proteins are the major cellular targets of \mOH-mediated radiation damage  \citep{Du:2004td}. Post-translational modification of proteins due to radical interactions can lead to a wide array of potentially lethal consequences, including formation of reactive peroxide species \citep{Davies:1995wz}, formation of insoluble aggregates \citep{Barnham:2004tl}, loss of enzymatic function, and destabilization of functional complexes; at minimum, such damage increases the rate of protein expression and controlled degradation required for homeostasis, thereby raising the metabolic cost of cellular survival. Although \mOH production is a threat to irradiated organisms in any environment, some environmental conditions may partially or substantially ameliorate it.  In particular, environments that favor the scavenging of \mOH by other chemical species reduce the level of chemical stress to which organisms are subject, and may thus allow for greater radiation tolerance.  Because many reactions involving \mOH are (or are near) diffusion-limited kinetics, the \mOH diffusion coefficient is of particular relevance to extremophile biochemistry in irradiated environments.

Among the environments of particular interest for novel biochemistry are those involving low temperatures and/or high pressures.  Such environments occur in deep ocean and ice/rock interfaces on Earth, and in subsurface oceans in the outer solar system.  Because the diffusion coefficient for \mOH at low temperature and high pressure has not been measured to date, we here estimate it using atomistic molecular dynamics (MD) simulations, employing a novel hierarchical Bayesian inference scheme to infer the diffusion coefficient while correcting for finite size effects.  In the process, we also parameterize a CHARMM-compatible model for \mOH for use with the TIP4P/2005 water model (chosen for its ease of implementation with standard MD platforms and its performance in reproducing properties of bulk water over a wide temperature and pressure range).  We summarize our posterior inference in the form of a simple log-log polynomial model that can be used to reproduce our simulation-based estimates of \DiOH over a range of temperatures and pressures.

The remainder of the paper is structured as follows:  Sec.~\ref{sec_methods} describes our procedures, in particular including the parameterization of the \mOH model (Sec.~\ref{sec_param}) and inference for the diffusion coefficient (Sec.~\ref{sec_DOH}).  Our results are summarized in Sec.~\ref{sec_results}, and Sec.~\ref{sec_conclusion} concludes the paper.

\section{Methods} \label{sec_methods}

Our interest is inferring the diffusion coefficient of the hydroxyl radical in aqueous solution, \DiOH, as a function of temperature and pressure.  We begin with parameterization of a model for \mOH in TIP4P/2005 water, followed by our simulation design.  We then describe our approach for inferring the diffusion coefficient, \DiOH, from simulated water and radical trajectories.  Results are shown in Sec.~\ref{sec_results}.

\subsection{Parameterization of the \mOH Model} \label{sec_param}

To perform atomistic simulations of \mOH in solution, we parameterize a CHARMM-based \citep{best.et.al:jctc:2012} model for \mOH in TIP4P/2005 water \citep{abascal.vega:jcp:2005}.  We employ TIP4P/2005 because of its strong performance in reproducing the diffusion constant of water over a wide range of temperatures and pressures \citep{tsimpanogiannis.et.al:ms:2019}.  Partial charges, bond length, mass, and force constant for \mOH are taken from \citet{pabis.et.al:pccp:2011}, who performed combined DFT and MD studies of \mOH in BJH water at physiological temperature and pressure \citep{bopp.et.al:cpl:1983}; unfortunately, the non-bonded interactions employed cannot be directly adapted to the CHARMM forcefield, and hence it is necessary to parameterize them directly.  The non-bonded interactions in question are defined by a Lennard-Jones potential of the form
\begin{equation*}
\begin{split}
\epsilon_{LJ}(R_{AB}|\epsilon_A,\epsilon_B,\RA,\RB) =\\ 
\sqrt{\epsilon_A \epsilon_B} \left[\left(\frac{\RA+\RB}{R_{AB}}\right)^{12} + 2\left(\frac{\RA+\RB}{R_{AB}}\right)^{6}\right],
\end{split}
\end{equation*}
where $R_{AB}$ is the distance between atoms $A$ and $B$, $\epsilon_A$ and $\epsilon_B$ are species-specific well-depth parameters, and \RA and \RB are ``half-radii'' that determine the zero-point of the interatomic force.  Here, we must determine these parameters for the two respective atoms of \mOH, given the TIP4P/2005 parameters (which we take as fixed). As our interest is in \DiOH, we optimize the \mOH non-bonded interaction parameters (\EO, \EH, \RO, and \RH) so as to reproduce the measured value of \DiOH=0.23\AA$^2$/ps for \mOH in water at 298K and atmospheric pressure \citep{dorfman.adams:bk:1973}, holding all other factors constant.

Our protocol proceeded as follows.  We began with a quasi-random search of the parameter space, drawing 250 points from the intervals $\EO,\EH\in (-0.5,0)$ kcal/mol, $\RO\in (0,3)$ \AA, and  $\RH\in (0,2)$ \AA{ } using a four-dimensional Halton sequence (bases 2, 3, 5, and 7).  For each parameter vector, a 1 ns atomistic simulation of one \mOH in TIP4P/2005 water under periodic boundary conditions was performed, with frames sampled every 0.5 ps (integrator step size 2 fs).  Simulations were initialized with a cubic box of $\approx 20$ \AA{ } side length at 1 atm and 298K; two adjustment phases of 100 ps each were performed (with box sizes adjusted for PME calculations after each phase) prior to the production run, with both adjustment and production phases performed with the $NpT$ ensemble.  Langevin dynamics with an interval of 1/ps were employed for temperature control, and a Langevin-Nos\'{e}-Hoover piston with a period of 100 ps was used to maintain constant pressure \citep{martyna.et.al:jcp:1994,feller.et.al:jcp:1995}.  Rigid bonds were maintained for all waters, with the O-H bond of \mOH left flexible.  All simulations were performed using NAMD \citep{phillips.et.al:jcc:2005}, with initial conditions created using VMD \citep{humphrey.et.al:jmg:1996}, psfgen \citep{ribeiro.et.al:sw:2020}, and Packmol \citep{martinez.et.al:jcc:2009}.  Each simulated trajectory was then unwrapped using the protocol of \citet{bulow.et.al:jcp:2020} to account for changing box sizes, and each frame was centered at its centroid to correct for net drift.  Molecular positions were extracted via the resulting oxygen atom coordinates.

To obtain initial estimates for \DiOH at each parameter value, the covariance-based estimator of \citet{bullerjahn.et.al:jcp:2020} was applied to each processed \mOH trajectory; this estimator is computationally efficient, and was found to work well in pilot runs using both \mOH and \mHTO under these simulation conditions.  These raw estimates were then corrected for finite sample sizes using the analytical correction factor of \citet{yeh.hummer:jpcB:2004},
\[
D_\infty = D_{\mathrm{MD}} + \frac{\zeta k_B T}{2 \pi \eta L},
\]
where $D_\infty$ is the diffusion constant at infinite size, $D_{\mathrm{MD}}$ is the diffusion constant under PBC obtained from MD simulation,  $\zeta \approx 2.8373$ is a numerical constant, $\eta$ is the shear viscosity of the solvent, and $L$ is the box length.  For $\eta$, the TIP4P/2005 viscosity of 8.55$\times 10^{-4}$ Js/m$^3$ from \citet{gonzalez.absacal:jcp:2010} was employed, and $L$ was taken to be the cube root of the mean box volume over the simulation.  The size-corrected estimates of \DiOH were then retained for further analysis.

To obtain an initial estimate of the non-bonded parameters, the ARE in the \DiOH estimate (i.e. $|0.23-\widehat{\DiOH}|/0.23$) was computed for each parameter vector.  Inspection of the resulting estimates revealed two regions of high performance (Fig.~\ref{f_ARE}).  To distinguish among the competing regions, energies were calculated for the interaction of a single \mOH and \mHTO in both conventional hydrogen bonding and ``flipped'' (i.e., unfavorable H-H or O-O configurations) over a range of 1-5 \AA{ } (Figs.\ref{f_parLJ}-\ref{f_parHO}).  The single region for which properly oriented hydrogen bonds were more favorable and the ARE was low was selected for further analysis.  (The most favorable point in this stage was $\EO \approx -0.224$, $\EH \approx -0.291$, $\RO \approx 0.802$, $\RH \approx 0.752$.)

Following this first stage of calibration, we performed a subsequent stage of refinement by repeating the procedure with search using larger simulated systems over a smaller parameter range; this secondarily confirmed that estimates were robust to system size.  As before, a four-dimensional Halton sequence was used to select parameters, with 25 draws taken over the range $\EO \in (-0.26,-0.18)$ kcal/mol, $\EH\in (-0.31,-0.27)$ kcal/mol, $\RO\in (0.6,1.1)$ \AA, and  $\RH\in (0.72,0.78)$ \AA. For each parameter vector, the above simulation and analysis protocol was followed, with the exception that initial box size was increased to 50 \AA.  Estimated size-corrected AREs were obtained for each parameter value, and the parameter vector yielding the minimum error was selected for final use.  (As above, energy calculations were employed to verify that the model favored the correct \mOH-\mHTO orientation.)

\begin{figure}
  \centering
  \includegraphics[width=0.5\textwidth]{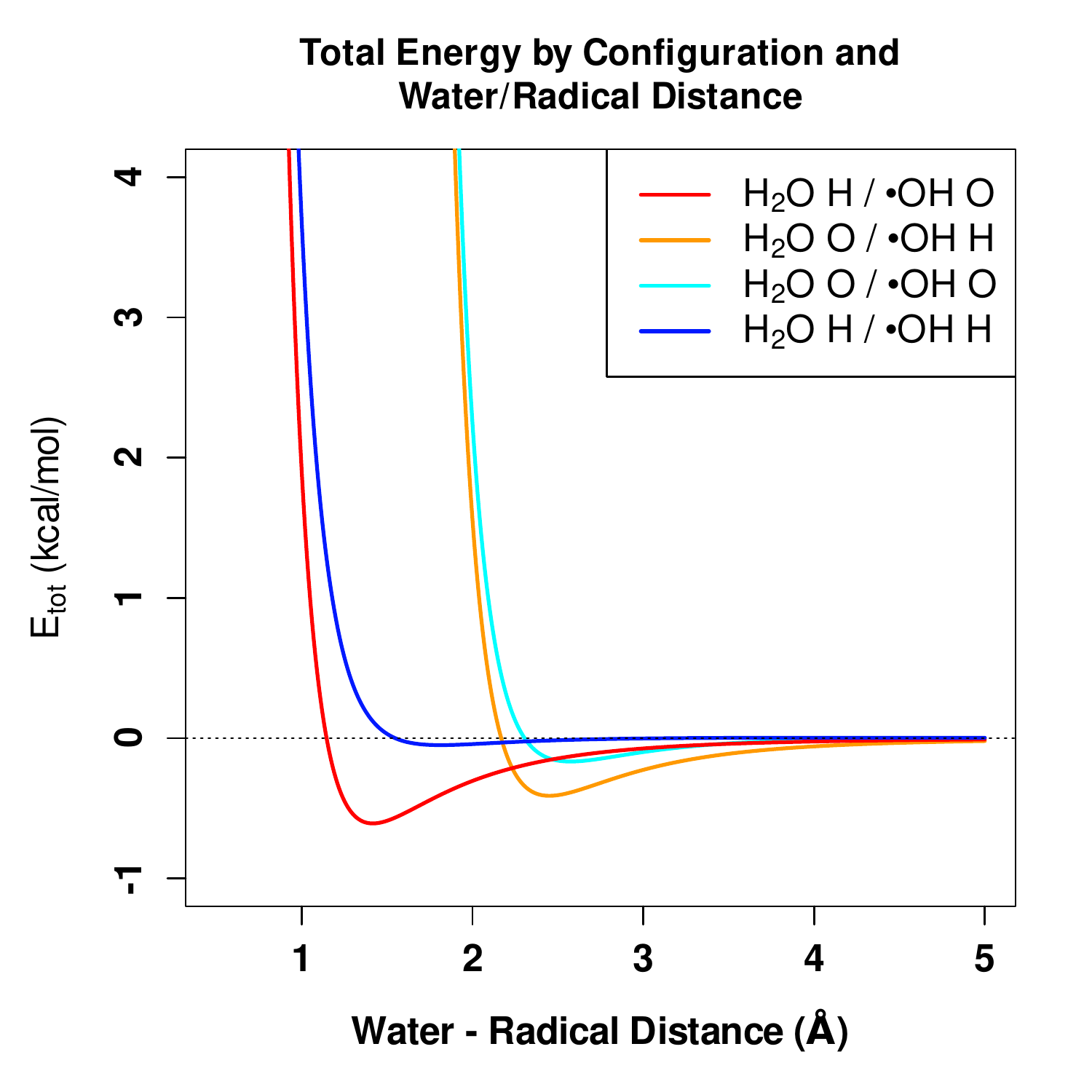}
  \caption{ Non-bonded interaction energy for \mHTO vs. \mOH as a function of configuration and distance between proximate atoms.  The final model correctly favors standard donor/acceptor orientations over ``flipped'' orientations, with the \mHTO donor interaction slightly favored over the \mOH donor interaction. \label{f_finHB}}
\end{figure}

The final parameters for the \mOH model (including both predetermined and calibrated parameters) are $r_{\mathrm{OH}}=0.9751$ \AA, $E_{\mathrm{OH}}=475.6$ kcal/mol/\AA$^2$, $m_\mathrm{O}=15.994$ Da, $m_\mathrm{O}=1.008$ Da, $\EO=-0.235$ kcal/mol, $\EH = -0.3068$ kcal/mol, $\RO = 0.78185$ \AA, and $\RH = 0.746939$ \AA.  The relatively small value of \RO seems to be necessary for proper interaction with the TIP4P/2005 water model, as larger values lead either to inappropriate donor/acceptor orientations for inaccurate values of \DiOH (Figs.~\ref{f_parOH},\ref{f_parHO}); the selected value leads to reasonable water/radical interactions, as shown in Fig.~\ref{f_finHB}.  The size-corrected diffusion constant estimate for the final model in the last selection round was $\widehat{\DiOH}\approx 0.222$ \AA$^2$/ps, for an ARE of approximately 4\% versus the target value of 0.23 \AA$^2$/ps.

\subsection{Diffusion Simulation}

Given the parameterized model, we simulate the diffusion of \mOH in TIP4P/2005 water at multiple temperatures and pressures.  The basic simulation strategy is equivalent to that described in Section~\ref{sec_param}, with the following modifications.  First, in order to ensure a high level of precision in our trajectory calculations, we sample trajectories every 0.25 ps instead of every 0.5 ps; additionally, we add an extra 100 ps to the pre-production run for each trajectory, which is not included in the 1 ns employed for estimation.  Rather than employing the covariance estimator for \DiOH, we use a Bayesian approach as described in Section~\ref{sec_inference}. And, finally, because the analytical correction for system size depends on the viscosity (which is not known for TIP4P/2005 over the range of temperatures and pressures studied here), we employ statistical corrections involving variable system sizes as explained below.

We perform simulations at 263, 273, 283, and 298K, and at 1, 10, 100, 1000, and 10000 atm pressure (full factorial design).  To allow statistical correction for finite size effects (and to reduce simulation-related error) we perform 30 replicate simulations for each condition, with initial box sizes evenly spaced from 20 to 50 \AA.  The unfolded trajectories from each size replicate are then used to infer the diffusion coefficient, as described below.

\subsection{Inference for the Diffusion Coefficient} \label{sec_DOH}

To infer \DiOH from simulation, we must account for both transients that affect the observed within-trajectory diffusion rate, and finite size effects that lead to systematic variation across trajectories for systems of differing size.  Here, we use a two-stage Bayesian inference strategy, first obtaining local posterior estimates of $D_{\mathrm{MD}}$ using a modified Brownian motion process model, and then integrating these local estimates via a hierarchical model that combines estimates of \DMDOH and \DMDHTO across systems of varying size to obtain final estimates of \DiOH.

\subsubsection{Local Estimation of $D_{\mathrm{MD}}$} \label{sec_localest}

Although the covariance estimator of \citet{bullerjahn.et.al:jcp:2020} is both computationally and statistically efficient at high temperatures, it performs less well when the diffusion coefficient becomes small relative to the the background noise for which it controls (see discussion in \citet{bullerjahn.et.al:jcp:2020}).  Here, we thus use a strategy of Bayesian estimation for the local (size uncorrected) diffusion constant $D_{\mathrm{MD}}$, which both makes more complete use of data and provides regularization of the resulting estimator.  The model employed here is based on the model (somewhat tacitly) underlying the generalized least squares estimators of \citet{bullerjahn.et.al:jcp:2020}, namely a latent Brownian motion process with a Gaussian observation mechanism.  Given regularly spaced observations $Y=(Y_1,Y_2,\ldots)$ at times $1, 2, \ldots$, the process may be defined in one dimension by
\begin{gather}
X_{t+1} = X_t + Z_t\\
Y_{t+1} = X_{t+1} + W_t,
\end{gather}
where $Z_t$ is iid $N(0,\sigma^2)$, and $W_t$ is iid $N(0,a^2)$.  Physically, $X$ here represents a ``true'' or idealized Brownian motion with independent perturbations given by $Z$, while $W$ reflects idiosyncratic noise factors arising from non-Brownian transients.  The diffusion constant corresponds to $D=\sigma^2/2$ (in the squared distance units of $Y$ divided by the time between steps). We may observe that this is a (discretely measured) Gaussian process with covariance function $K(Y_i,Y_j)=\min(i,j)\sigma^2+a^2 I(i=j)$, and hence the likelihood is given by (conditioning on and centering the first observation)
\begin{equation}
p(Y=y|\sigma^2,a^2) = MVN(y|\mathbf{0},K(y)),
\end{equation}
where $\mathbf{0}$ is the 0-vector, and $K(y)$ is the Gram matrix of the observed sequence.  This is straightforward to work with, although computationally expensive when the number of time points becomes large, and can be pooled across dimensions in the isotropic case (as is done here). 

To define priors on the variance parameters, we first observe that on our physical scale of interest (\AA$^2$/ps) and over the range of conditions considered here, it is \emph{a priori} unlikely that $\sigma^2$ will exceed 0.5; likewise \citep[per][]{bullerjahn.et.al:jcp:2020} it is reasonable to expect $a^2$ to be comparable to (or possibly smaller than) $\sigma^2$.  We thus use independent half-Gaussian priors for $\sigma^2$ and $a^2$, with a scale of 0.5, which is relatively flat over the region of interest while discouraging strongly unphysical values.  (Note that this is equivalent to $L_2$ regularization of the variance parameters.)  

For parameter estimation, it is natural here due to both the Gaussian structure of the problem and the large data size to employ MAP estimation, invoking the Laplace approximation \citep{gelman.et.al:bk:2003} to obtain posterior standard deviations.  We perform estimation using a custom R \citep{rteam:sw:2021} script, with direct optimization of the log posterior surface using BFGS \citep{nash:bk:1990}; the \texttt{mclust} package \citep{scrucca.et.al:rj:2016} was used for efficient calculation of the multivariate Gaussian log-likelihood.  The posterior standard deviation of $\hat{D}$ was obtained via the Hessian of the negative log posterior about the posterior mode (exploiting the linear relationship between $D$ and $\sigma^2$).  Due to the cost of computing the Gram matrix, the initial unfolded and drift-corrected trajectories were downsampled from 0.25 ps to 0.5 ps resolution, and split into two segments of 500 ps length (i.e., 1000 observations); these were pooled in the likelihood calculation.  In the case of \mHTO trajectories, trajectories for all water molecules were pooled and jointly analyzed.  This process led to estimates of the posterior mode (assumed equal to the mean, under the Laplace approximation) and standard deviation for $D_{\mathrm{MD}}$ for both \mOH and \mHTO under each condition and at each box size.  These estimates were then integrated to estimate \DiHTO and \DiOH in each condition as described below.

\subsubsection{Estimation of \DiOH} \label{sec_inference}

\begin{figure}
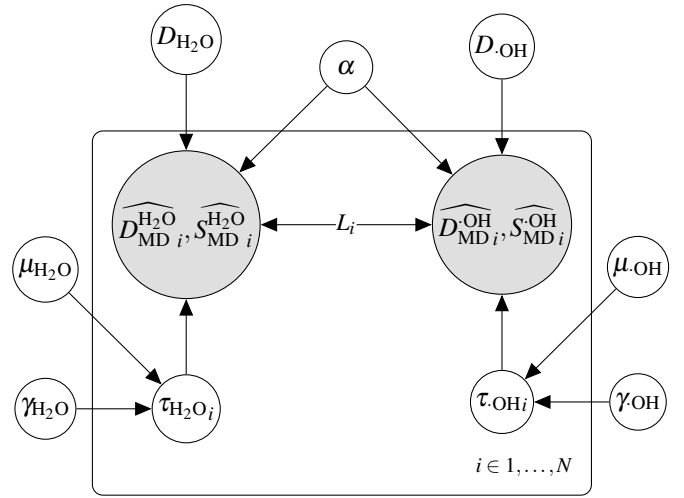

\centering
\tikz{ %
  \node[const] (Li) {$L_i$} ; %
  \node[latent, above=of Li, yshift=0.6cm] (alpha) {$\alpha$} ; %
  \node[obs, left=of Li] (DmdH2Oi) {$\widehat{\DMDHTO}_i,\widehat{\SMDHTO}_i$} ; %
  \node[obs, right=of Li] (DmdOHi) {$\widehat{\DMDOH}_i,\widehat{\SMDOH}_i$} ; %
  \node[latent, above=of DmdH2Oi] (DH2O) {\DiHTO} ; %
  \node[latent, above=of DmdOHi] (DOH) {\DiOH} ; %
  \node[latent, below=of DmdH2Oi] (eVh2oi) {${\tauHTO}_i$} ; %
  \node[latent, below=of DmdOHi] (eVohri) {${\tauOH}_i$} ; %
  \node[latent, left=of eVh2oi] (evScah2o) {\gammaHTO} ; %
  \node[latent, right=of eVohri] (evScaohr) {\gammaOH} ; %
  \node[latent, above=of evScah2o] (evLoch2o) {\muHTO} ; %
  \node[latent, above=of evScaohr] (evLocohr) {\muOH} ; %
  \plate[inner sep=0.25cm] {plate} {(DmdOHi) (DmdH2Oi) (eVohri) (eVh2oi) (Li) } {$i \in 1,\ldots,N$}; %
  \edge {DH2O,alpha,Li,eVh2oi} {DmdH2Oi} ; %
  \edge {DOH,alpha,Li,eVohri} {DmdOHi} ; %
  \edge {evLoch2o,evScah2o} {eVh2oi} ; %
  \edge {evLocohr,evScaohr} {eVohri} ; %
}
\caption{\label{f_mod} Structure of the diffusion constant model.  White circles indicate latent quantities, while observed random quantities are indicated by shaded circles; fixed values are uncircled.  Elements within the central plate are replicated $N$ times, while those outside are pooled.}
\end{figure}

Estimation of small-molecule diffusion constants is challenging due both to the need to correct for finite-size effects and a high level of idiosyncratic variation between trajectories that is difficult to account for; moreover, a single molecule trajectory provides relatively little information per simulation run (as opposed to the large number of solvent molecule trajectories obtained on each run).  Here, we address both issues via a hierarchical Bayesian model that pools information between \mHTO and \mOH trajectories, and that incorporates multiple sources of variation.  The model (whose structure is described pictorially in the plate diagram of Figure~\ref{f_mod}) is defined as follows.
 
We begin with the observation that, if \DiOH and \DiHTO are the respective bulk diffusion constants for \mOH and \mHTO, then
\begin{gather}
\DiOH = {\DMDOH}_i - \alpha/L_i\\
\DiHTO = {\DMDHTO}_i - \alpha/L_i,
\end{gather}
where ${\DMDOH}_i$ and ${\DMDHTO}_i$ are the diffusion coefficients for a PBC system with length scale $L_i$, and $\alpha$ is the system size scaling coefficient.  Of these, only $L$ is observed.  We do, however, have local estimates of ${\DMDOH}_i$ and ${\DMDHTO}_i$, which we model as
\begin{gather}
\widehat{{\DMDOH}_i} = {\DMDOH}_i + E_{\mOH}\\
\widehat{{\DMDHTO}_i} = {\DMDHTO}_i + E_{\mHTO},
\end{gather}
where  $E_{\mOH} \sim N(0,\widehat{\SMDOH}_i+{\tauOH}^2_i)$ and $E_{\mHTO} \sim N(0,\widehat{\SMDHTO}_i+{\tauHTO}^2_i)$ represent deviations from the idealized local diffusion coefficients.  The error variance is modeled via two components: the posterior variance from the local model of Section~\ref{sec_localest} ($\widehat{\SMDOH}_i,\widehat{\SMDHTO}_i$); and the excess variances ${\tauOH}^2_i$ and ${\tauHTO}^2_i$ representing trajectory-specific idiosyncratic deviations not reflected by the within-trajectory estimates.  We take the square roots of the excess variances to be generated by 0-truncated normal distributions, i.e. $\tauOH \sim TN_0(\muOH,\gammaOH^2), \tauHTO \sim TN_0(\muHTO,\gammaHTO^2)$, with weakly informative standard half-Cauchy priors on the $\mu_{*}$ and $\gamma_{*}$ parameters.  We observe that this prior structure can be seen as flexibly generalizing several standard regression-like models: in the limit as $E_{*}\to 0$, we recover a model akin to weighted least squares, with weights based on the locally estimated variances; when $E_{*}\gg\hat{S_{*}}_i$ but $\gamma_{*} \to 0$, recover a model akin to a standard homoskedastic regression; and when $\gamma_{*} \gg 0$ we obtain a robust regression with a heavy-tailed error distribution.  Finally, we take $\alpha$ to be \emph{a priori} uniform on (0,0.75) (as the 0.75 is expected to be strictly larger than the value of $\alpha$ for the conditions studied here).

Given the above, we perform posterior simulation using the No-U-Turn Hamiltonian Monte Carlo algorithm \citep{homan.gelman:jmlr:2014} from the Stan library \citep{rstan:sw:2020,stan:sw:2020}.  4 chains were employed for each condition, with $10^5$ burn-in iterations per chain followed by $10^5$ additional iterations from which 1000 were retained (i.e., a thinning interval of 100) for a final sample size of 4000 draws per condition.  Convergence was assessed with $\hat{R}$ \citep{gelman.rubin:ss:1992}. Posterior means and 95\% posterior intervals were obtained for \DiOH and \DiHTO for each condition for subsequent analysis, as discussed below.

\section{Results and Discussion} \label{sec_results}

\begin{figure*}
  \centering
  \includegraphics[width=0.9\textwidth]{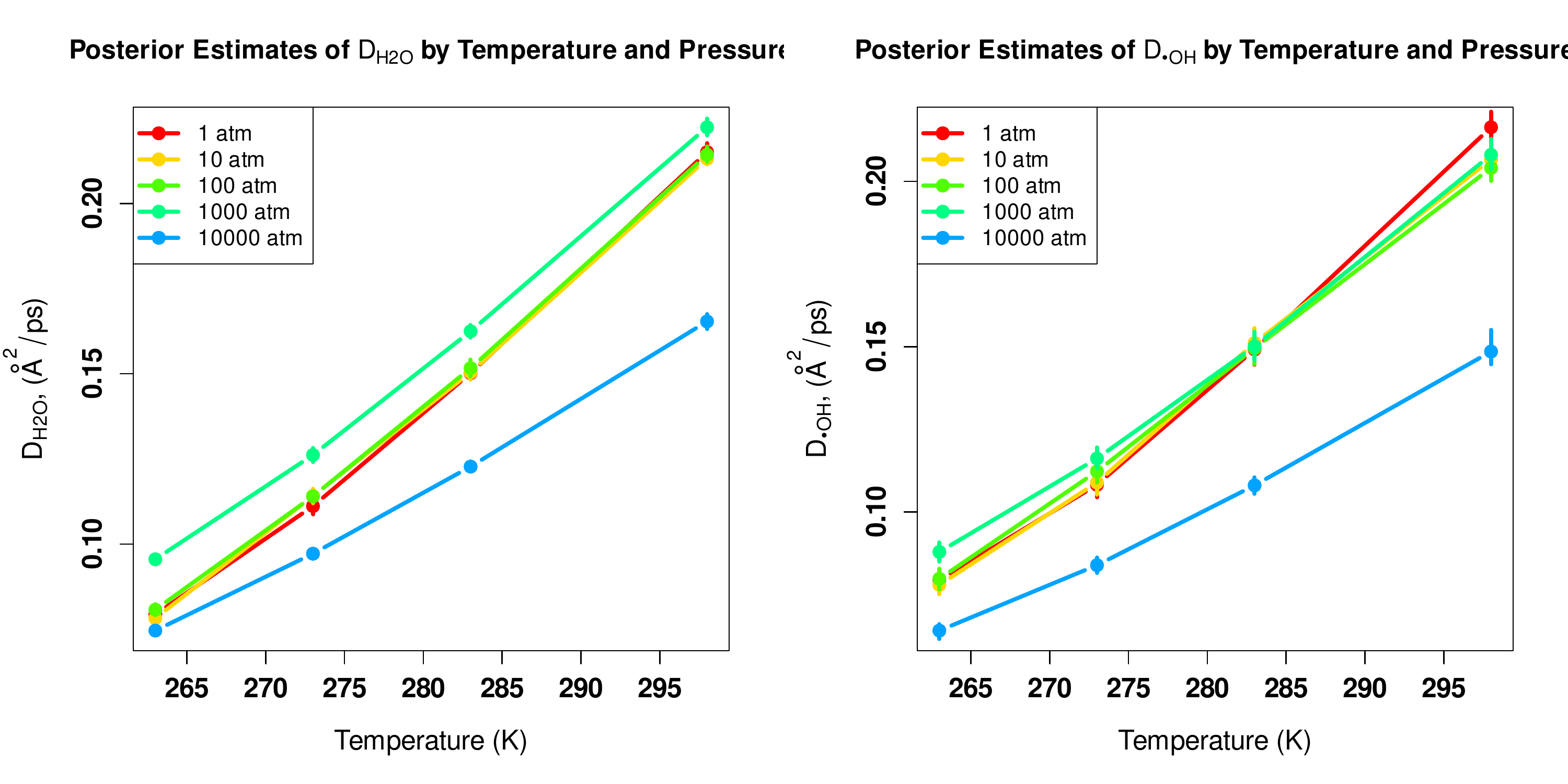}
  \caption{Posterior mean estimates and 95\% posterior intervals for \DiHTO and \DiOH. (In some cases, interval width is narrower than point sizes.)  Both behave similarly at low and very pressures, but \DiOH lacks the accelerated diffusion seen in \mHTO at moderately high pressure. \label{f_DH2O_DOH}}
\end{figure*}

Fig.~\ref{f_DH2O_DOH} shows the estimated bulk diffusion coefficients for \mOH and \mHTO at the simulated temperatures and pressures.  Although \DiOH and \DiHTO approximately coincide under ambient conditions, we observe some differences in their response to temperature and pressure.  While TIP4P/2005 reproduces the experimentally observed increase in \mHTO diffusion rate under moderate pressure, \mOH diffusion slows with pressure at moderate temperatures (eventually gaining very modest acceleration in the low temperature regime).  At pressures approaching 10000 atm, both water and the hydroxyl radical slow considerably (as expected from the well-known increase in water viscosity in this regime), with \DiOH diffusing more slowly than water despite its smaller size.  This difference in the high-pressure behavior of \mOH versus its behavior under ambient conditions may be related to the observation of an MD study in BJH water at relatively high temperature (310K) and 1 atm that \mOH tends to occupy cavities in the \mHTO hydrogen bonding network \citep{pabis.et.al:pccp:2011}.  At high pressure, occupancy of such cavities will be highly favorable, and may retard diffusion.

Changes in \mOH diffusion may have implications for radiochemistry in high pressure/low temperature environments.  The direct products of $\gamma$-irradiation of water are formed in ``spurs,'' localized regions of high concentration for these highly reactive species (most directly, \mOH, \radH, and hydrated electrons) \citep{Schwarz:1969vg, Parajon:2008vs}. When a $\gamma$-ray interacts with a water molecule, it generates a highly reactive excited state,

\begin{equation}
\mHTO + \gamma \rightarrow \mHTO*
\end{equation}

which can either directly decay to form a hydrogen atom and a hydroxyl radical,

\begin{equation}
\mHTO* \rightarrow \radH + \mOH
\end{equation}

or it can lose an electron to the surrounding solution, resulting in a radical cation that in turn decomposes to yield a hydroxyl radical and a proton:

\begin{equation}
\mHTO* \rightarrow \posHTO + \e
\end{equation}

\begin{equation}
\posHTO  \rightarrow \mH + \mOH
\end{equation}

These reactions serve as the starting point for the formation of a complex mixture of reactive oxygen species \cite{Ershov:2008uz} and/or reactions of these primary products with biomolecules \citep{Omar:2021vc}. The recent advent of experimental techniques enabling direct detection of the long-hypothesized \posHTO intermediate \citep{Loh:2020te} and observation of attosecond dynamics in liquid water \citep{Jordan:2020wm}, as well as the discovery of new reaction pathways \citep{Thurmer:2013wj, Ren:2018wj} have led to renewed interest in the fundamental radiochemistry of aqueous solutions. Analysis of the relevant reactions requires reasonable estimates of the diffusion coefficients of key chemical species, as small solutes can undergo anomalous diffusion in water \cite{Kirchner:2002wf, Roberts:2009tp, Marx:2010vs}, particularly under extreme conditions.   \mOH, the focal species of this work, has been the subject of previous studies; however, these have generally focused on high temperature, e.g. \citep{Plugatyr:2005tv}.  The different patterns of \mOH diffusion seen in low-temperature, high-pressure regime suggest considerable value in further experimental studies of these environments.

\begin{figure}
  \centering
  \includegraphics[width=0.5\textwidth]{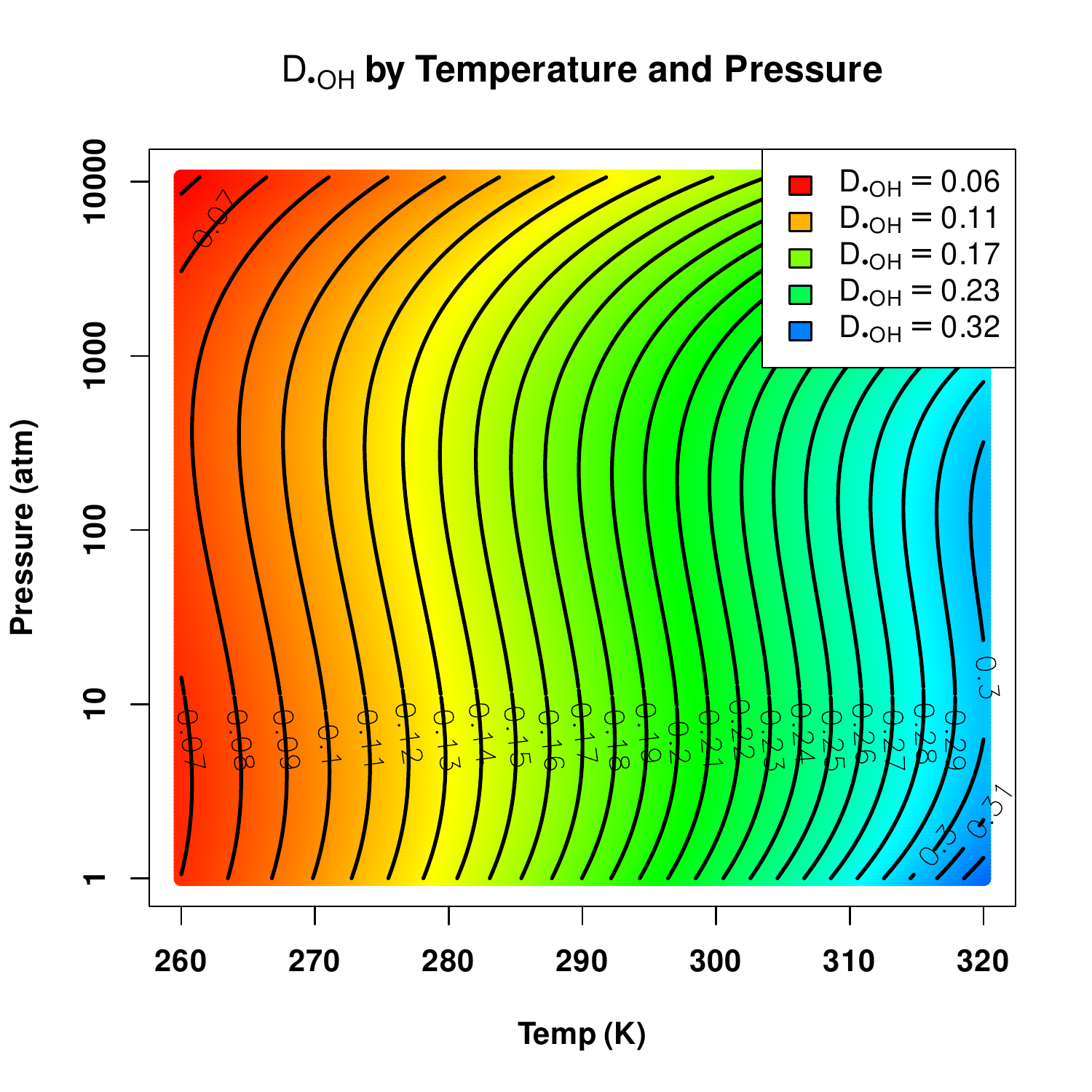}
  \caption{$\widehat{\DiOH}$ as a function of temperature and pressure, based on posterior inference from simulations in TIP4P/2005 water.  \mOH diffusion is slightly faster at moderate pressures when temperatures are low (though less so than \mHTO), but slows considerably at high pressures.  \label{f_DOH}}
\end{figure}

To better summarize the posterior surface of \DiOH in response to temperature and pressure, we fit a least-squares log-log polynomial approximation to the posterior means.  The resulting summary model ($R^2=0.995$, standard error 0.031 on 13 residual degrees of freedom (log scale), RMSE $0.0035 \unit{\square\angstrom\per\pico\second}$ on the phenomenological scale) is given by
\begin{equation}
\begin{split}
\widehat{\DiOH} = \exp\left[\num{-4.410203e+02} + \num{1.476812e+02}  T' \right.\\
\left. + \num{8.225372e-01} P' + \num{-1.238130e+01} T'^2 \right.\\
\left.+ \num{2.814801e-02} P'^2 + \num{-1.582325e-01} T' P' \right.\\
\left. + \num{-2.622113e-03} P'^3\right] \unit{\square\angstrom\per\pico\second} 
\end{split}\label{eq_DOH}
\end{equation} 
where $T'$ and $P'$ are the dimensionless quantities $T'=\log \left(T \; \unit{\per\kelvin}\right)$ and $P' = \log\left(P \; \unit{\per\atm}\right)$.  The resulting surface is shown in Fig.~\ref{f_DOH}.  As expected from the known coincidence of \DiOH and \DiHTO under ambient conditions, \DiOH has broadly similar behavior to \DiHTO.  However, \mOH shows less accelerated diffusion at moderate pressures, and slows considerably at very high pressures (even in excess of \mHTO).

\section{Conclusion}\label{sec_conclusion}

Diffusion rates for the hydroxyl radical differ substantially at low temperatures and high pressures.  For the most part, the pattern of variation in \DiOH qualitatively follows that of \DiHTO (as would be expected for a small molecule in aqueous solution), but we do not observe the same extent of enhanced diffusion at moderate pressures (particularly at higher temperatures), and diffusion slows more substantially as pressures approach 10000 atm.  This may have an impact on the efficiency of radical-scavenging mechanisms in high-pressure terrestrial environments, and hypothetically in the even higher pressure conditions predicted to obtain within some outer solar system oceans.  

To estimate the \mOH diffusion coefficient, we combined atomistic molecular dynamics with hierarchical Bayesian inference that allows us to easily pool information between solvent and radical, and across multiple trajectories of varying size.  The modular structure of Bayesian models lends itself naturally to this application, as does the ease with which one can e.g. leverage \emph{a priori} information on model parameters or account for variance across trajectories.  This strategy (which builds on other recent work on statistical inference for the diffusion coefficient \citep{bullerjahn.et.al:jcp:2020}) is readily adapted to the estimation of diffusion coefficients for other systems, or to other, related physical quantities.

\begin{acknowledgments}
This work was supported by NASA award 80NSSC20K0620 to R.W.M. and C.T.B. We thank George Miller for helpful discussions about the chemistry of water radiolysis. 
\end{acknowledgments}

\section*{Data Availability Statement}

Data sharing not applicable – no new data generated.  CHARMM-compatible parameters for the \mOH model are included in the main text.

\bibliography{../bibhome/ctb,../bibhome/radiation}

\clearpage

\begin{widetext}
\begin{center}
{$\;$}\\ \vspace{0.5in}
\textbf{\large Supplemental Materials for \emph{Bayesian Estimation of the Hydroxyl Radical Diffusion Coefficient at Low Temperature and High Pressure from Atomistic Molecular Dynamics}}\\
\vspace{0.5in}
{\large Carter T. Butts and Rachel W. Martin\\
University of California, Irvine\\
7/28/21\\} \vspace{5in}
\end{center}
\end{widetext}

\setcounter{equation}{0}
\setcounter{figure}{0}
\setcounter{table}{0}
\setcounter{page}{1}
\makeatletter
\renewcommand{\theequation}{S\arabic{equation}}
\renewcommand{\thefigure}{S\arabic{figure}}
\renewcommand{\bibnumfmt}[1]{[S#1]}
\renewcommand{\citenumfont}[1]{S#1}

\clearpage
\begin{figure*}
  \centering
  \includegraphics[width=\textwidth]{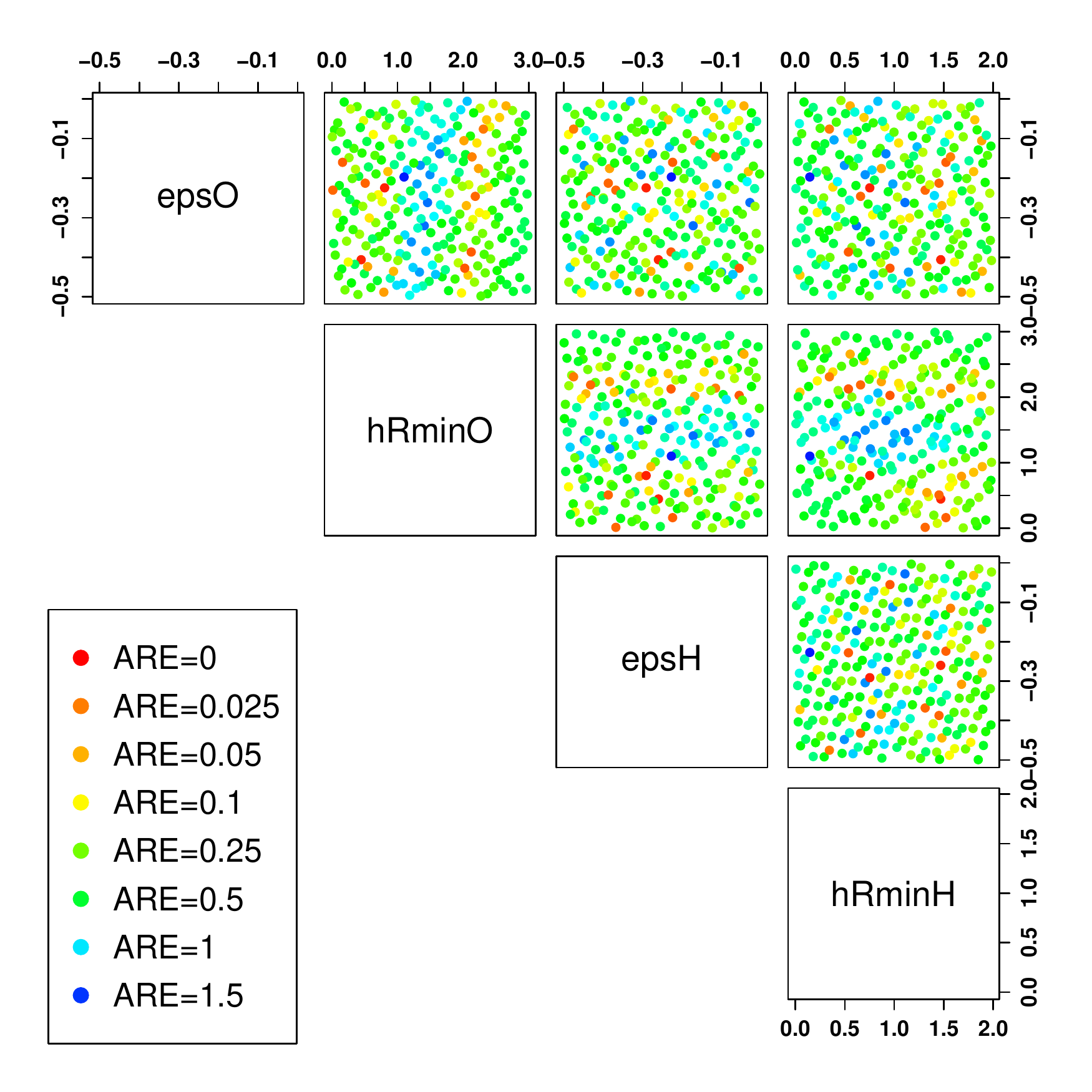}
  \caption{AREs for estimated \DiOH, 20 \AA box size simulations, by parameter value.  Parameters are indicated as follows: epsO=\EO, hRminO=\RO, epsH=\EH, hRminH=\RH. \RO is a critical parameter, with values between apx 1.3-1.8 \AA leading to especially poor diffusion performance.  Two broad regions with high-performing parameters are visible, motivating selection by examination of other aspects of model behavior. \label{f_ARE}}
\end{figure*}

\begin{figure*}
  \centering
  \includegraphics[width=\textwidth]{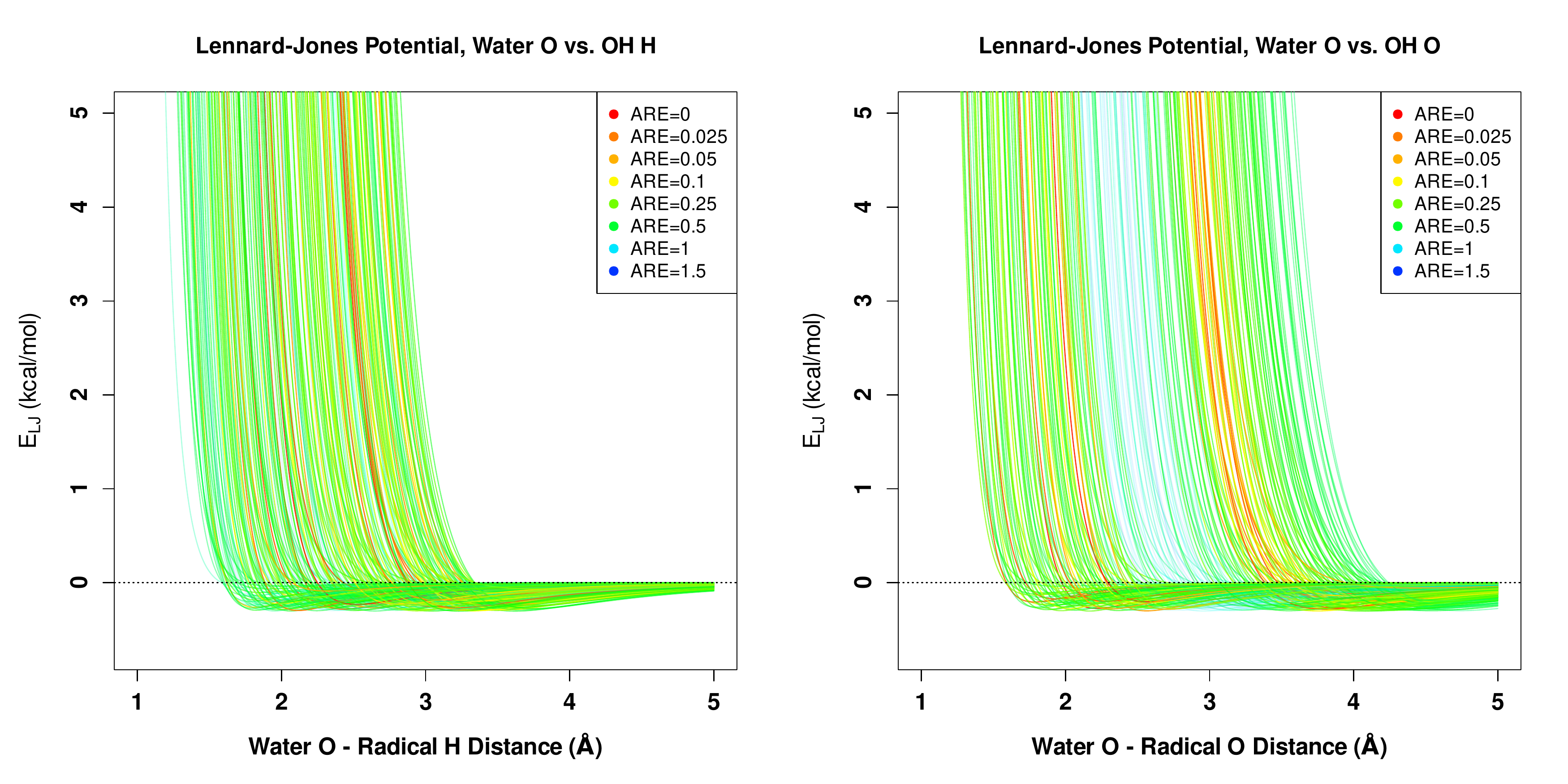}
  \caption{Lennard-Jones energy for interaction of water oxygen with \mOH atoms (as TIP4P/2005 includes no LJ effects for water protons, these are not included, and there are no orientation effects). Diffusion-accurate parameter values are characterized by an exclusion of attractive O-O interactions within a particular distance band, but have no specific relationship to the O-H distance.  \label{f_parLJ}}
\end{figure*}

\begin{figure*}
  \centering
  \includegraphics[width=\textwidth]{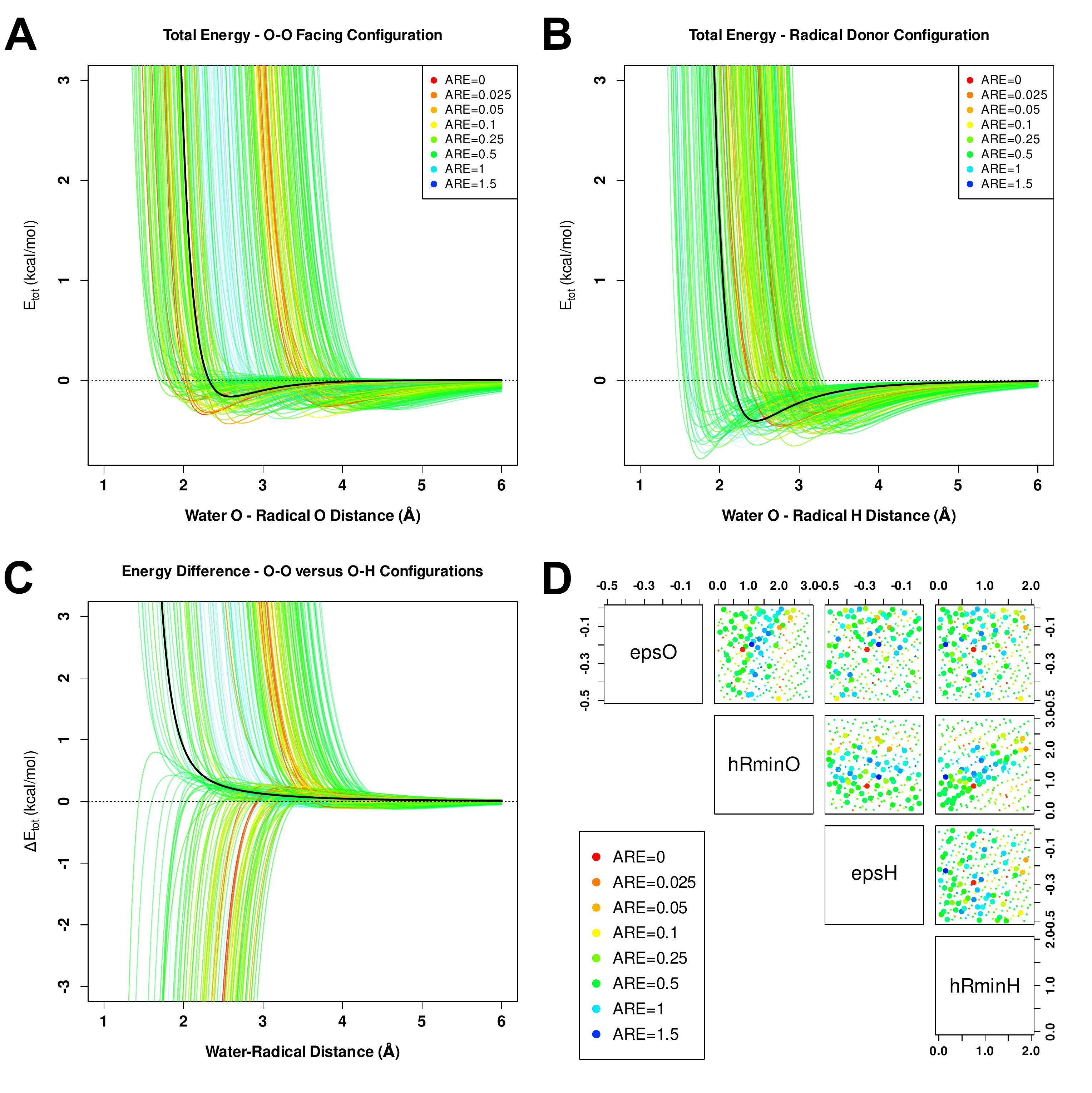}
  \caption{Winnowing of parameter vectors by hydrogen bonding behavior with \mHTO as acceptor; curves indicate energies for specific parameter vectors, with the vector selected in the initial round of optimization indicated in black.  (A) total \mHTO, \mOH interaction energy in an O-O facing configuration by distance.  (B) equivalent energies for the O-H configuration (\mHTO donor).  (C) difference between O-O and O-H energies by trajectory (positive values indicate that the O-O orientation is correctly unfavorable at the specified distance); note that, while some low-ARE parameters select the incorrect orientation, the chosen parameter vector favors the correct state.  (D) AREs by parameter (epsO=\EO, hRminO=\RO, epsH=\EH, hRminH=\RH), with points scaled by extent to which the corresponding parameter factors the correct versus incorrect orientation.  Restricting consideration to parameters with correct hydrogen bonding behavior greatly narrows the range of viable parameters. \label{f_parOH}}
\end{figure*}

\begin{figure*}
  \centering
  \includegraphics[width=\textwidth]{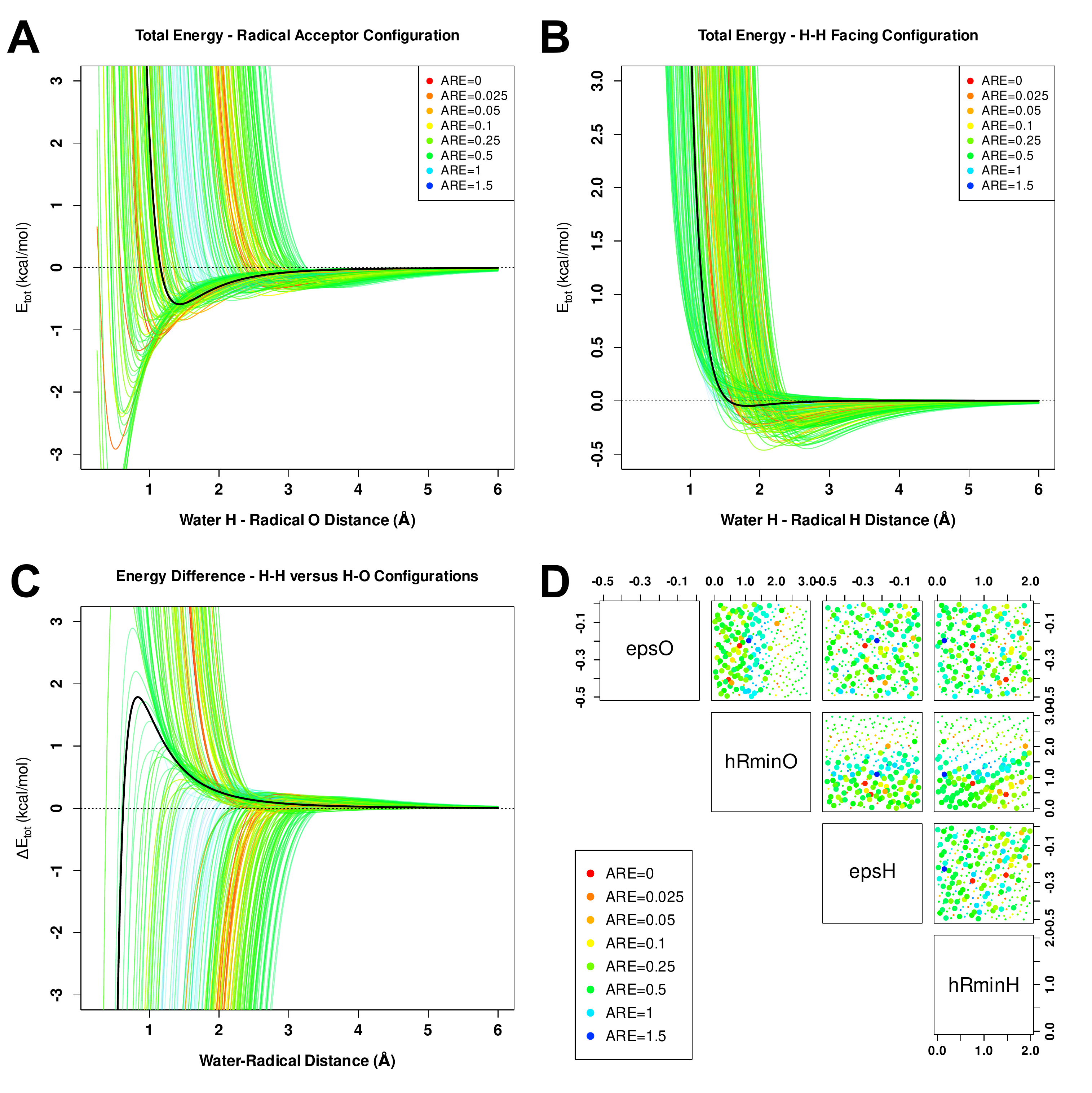}
  \caption{Winnowing of parameter vectors by hydrogen bonding behavior with \mHTO as donor; curves indicate energies for specific parameter vectors, with the vector selected in the initial round of optimization indicated in black.  (A) total \mHTO, \mOH interaction energy in an H-O facing configuration by distance (\mHTO donor).  (B) equivalent energies for the H-H configuration).  (C) difference between H-H and O-H energies by trajectory (positive values indicate that the H-H orientation is correctly unfavorable at the specified distance); the chosen parameter vector favors the correct state except at unphysically short radii (at which neither configuration is favorable).  (D) AREs by parameter (epsO=\EO, hRminO=\RO, epsH=\EH, hRminH=\RH), with points scaled by extent to which the corresponding parameter factors the correct versus incorrect orientation.  Restricting consideration to parameters with correct hydrogen bonding behavior narrows the range of viable parameters, but less so than the \mHTO acceptor case. \label{f_parHO}}
\end{figure*}

\end{document}